\documentclass[epj]{svjour}
\usepackage{graphics}

 \usepackage{epsf}
 \usepackage{epsfig}
 \usepackage{epic} 
 \usepackage{eepic} 
 \usepackage{rotating}

 \usepackage{amssymb}
 \usepackage{amsmath}

  \def \bm #1{ \mbox {\boldmath $#1$} }
\begin{document}
\title{Probing the deuteron structure at small $N - N$ distances by
       cumulative pion production}
\author{A.Yu.~Illarionov
 \and A.G.~Litvinenko \and G.I.~Lykasov
}                     
%
%
\institute{Joint Institute for Nuclear Research, 
           141980 Dubna, Moscow Region, Russia\\
\email{Alexei.Illarionov@jinr.ru}
}

\date{Received: date / Revised version: \today}
%
\abstract{
  The fragmentation of deuterons into pions emitted forward in the
  kinematic region forbidden for free nucleon-nucleon collisions
  is analyzed. It is shown that the inclusion of the non-nucleonic
  degrees of freedom in a deuteron results in a satisfactory
  description of the data for the inclusive pion spectrum and improves
  the description of the data about $T_{20}$. According to the data,
  $T_{20}$ has very small positive values, less than $0.2$, which
  contradicts the theoretical calculations ignoring these degrees of
  freedom.
} 
\maketitle

 The investigation of polarization phenomena by deu\-te\-ron fragmentation
 at intermediate and high energies in the kinematic region forbidden for
 hadron emission by free $N - N$ scattering has recently become very
 topical. These are the so-called cumulative processes.

 Cumulative proton production in the collision of polarized deuterons with
 the target results in information about the deuteron spin structure at
 small inter-nuclear distances. This can be seen from the experimental
 and theoretical study of deuteron fragmentation into protons at a zero
 angle (see \cite{Lykasov:1993} and references therein). The theoretical
 analysis of this reaction has shown that the tensor analyzing power
 $T_{20}$ and the polarization transfer coefficient $\kappa$ are more
 sensitive to the deuteron wave function (DWF), particularly to the
 reaction mechanism, than the inclusive spectrum. At present, not a single
 DWF relativistic form can describe $T_{20}$ measured by $D p \to p X$
 stripping at light-cone variable $x \ge 1.7$. On the other hand, the
 inclusion of the reaction mechanism, namely the impulse approximation and 
the secondary interaction of the produced hadrons, can describe both the
 inclusive spectrum and $T_{20}$ at $x \le 1.7$ using only the nucleonic
 degrees of freedom \cite{Lykasov:1993}. Among other things this phenomenon
 can be due to the fact that the deuteron structure at a high
 ($>~0.20$~GeV/\textit{c}) internal momentum (short inter-nuclear distances
 $<~1$~fm) is determined by non-nucleonic degrees of freedom. The inclusion
 of non-nucleonic degrees of freedom, (it can be a six-quark state,
 $\Delta\Delta$, $NN^\star$, $NN\pi$ and other states in the deuteron)
 allowed one to describe the data on the inclusive proton spectrum at
 $x \geq 1.7$ \cite{Lykasov:1993}.

 If we try to study the manifestation of non-nucleonic degrees of freedom,
 it is natural to investigate the cumulative production of different hadrons
 having different quark contents. Interesting experimental
 data on $T_{20}$ in the reaction $D p \to \pi X$, where the pion is
 emitted forward, have been published recently \cite{Afanasev:1998}.
 These data are presented as a function of the so-called cumulative scaling
 variable $x_{\mathcal C}$ (``cumulative number'') \cite{Stavinsky:1979}.
 (The value of $x_{\mathcal C}$ corresponds to the minimum mass,
 in nucleon mass units, of the part of the projectile nucleus (deuteron)
 involved in the reaction. Values of $x_{\mathcal C} \sim x$ larger than $1$
 correspond to cumulative pions.) It was found very small, approximately
 constant, value of the tensor analyzing power $T_{20}$ for the deuteron
 fragmentation into pions $D p \to \pi X$ at $x_{\mathcal C} \geq 1$.

 In a recent papers \cite{Illarionov:2002:EPJA,Illarionov:2001:ch} we have
 investigated the reaction of deuteron fragmentation into pions within
 framework of the relativistic impulse approximation. The mechanism of this
 reaction is mainly an impulse approximation as the secondary interaction
 or the final state interaction is very small and can be neglected
 \cite{Amelin:1981}. The main goal was to describe this reaction in a
 consistent relativistic approach using a nucleon model of the deuteron.
 A fully covariant expression for all quantities was obtained
 within the Bethe-Salpeter formalism. In this way we have obtained general
 conclusions about the amplitude of the process, which can not be drawn in
 the non-relativistic approach. The main investigation results can be
 summarized as follows:
 (I) It was evident from the behavior of the inclusive pion spectrum and
 particularly the tensor analyzing power $T_{20}$ at large $x_{\mathcal C}$,
 that the relativistic effects are sizeable. However, the state of theory is
 such that the unique procedure to include relativistic effects in the
 deuteron has not been found yet. An extreme sensitivity to different
 methods of the relativistic deuteron wave function was found for $T_{20}$
 at $x_{\mathcal C} \geq 1$. It was shows that the inclusion of the
 ${\mathcal P}$ wave contribution to the DWF within the Bethe-Salpeter
 approaches \cite{Gilman:2001,Umnikov:1997} results in a better
 (but not satisfactory) description of the data over the cumulative region.
 (II) It was demonstrated a large sensitivity of the inclusive spectrum of
 pions to the vertex of the $NN \to \pi X$ subprocess.
 In contrast to this, small sensitivity of $T_{20}$ to this vertex was
 found. This polarization observable is very sensitive to the DWF form,
 that can be used for systematic investigation of the DWF.
 (III) For the deuteron fragmentation into protons emitted forward,
 the tensor analyzing power $T_{20}$ is not described by standard
 nuclear physics using the nucleonic degrees of freedom at
 $x_{\mathcal C} \geq 1.7$ \cite{Lykasov:1993}.
 On the contrary, $T_{20}$ for the fragmentation $D p \to \pi X$
 cannot be described within the same assumptions over all region
 $x_{\mathcal C} \geq 1$ \cite{Illarionov:2002:EPJA}.

 In this paper we try to include the non-nucleonic degrees of freedom
 within the approach suggested in \cite{Lykasov:1993,Efremov:1988},
 the use of which has reproduced the data for the proton spectrum in the
 deuteron stripping rather well.

 We consider the inclusive reaction of deuteron fragmentation into a
 pion, $ \overrightarrow{D} + p \to \pi(0^\mathrm{o}) + X $,
 for the polarized deuteron and pion emitted to forward at initial
 energies of order few GeV.
 If the initial deuteron is only tensor aligned due to their $p_D^{ZZ}$
 component,  the inclusive spectrum of this reaction can be written in
 the form:
\begin{equation}
 \rho_{pD}^\pi \left(p_D^{ZZ}\right)=
 \rho_{pD}^\pi \left[1 + {\mathrm A}_{ZZ} \; p_D^{ZZ}\right] \, ,
\label{ND8}
\end{equation}
 where
 $\rho_{pD}^\pi \equiv \varepsilon_\pi \cdot {d\sigma_{pD}^\pi / d^3p_\pi}$
 is the inclusive spectrum for the case of unpolarized deuterons and
 ${\mathrm A}_{ZZ} \equiv \sqrt2T_{20}$
 $(-\sqrt2 \leq T_{20} \leq 1/\sqrt2)$ is the tensor analyzing power.
 In the relativistic impulse approximation they can be written in a fully
 covariant manner within the Bethe-Salpeter formalism
 \cite{Illarionov:2002:EPJA}:
\begin{align}
 \rho_{pD}^\pi &= \dfrac{1}{(2\pi)^3} \int
 \dfrac{\sqrt{\lambda(p,n)}}{\sqrt{\lambda(p,D)}} \left[
 \rho_{pN}^\pi\cdot\Phi^{(u)}(|\vec q|)\right]
 \dfrac{m^2d^3q}{E_{\vec q}} \, ;
\label{ND9} \\
 \rho_{pD}^\pi \,& {\rm A}_{ZZ} = - \dfrac{1}{(2\pi)^3} \int
 \dfrac{\sqrt{\lambda(p,n)}}{\sqrt{\lambda(p,D)}}
\nonumber \\
 &\times \left[\rho_{pN}^\pi\cdot\Phi^{(t)}(|\vec q|)\right]
 \left(\dfrac{3\cos^2\vartheta_{\vec q}-1}{2}\right)
 \dfrac{m^2d^3q}{E_{\vec q}} \, ,
\label{ND10}
\end{align}
 where $\lambda(p_1, p_2) \equiv (p_1p_2)^2 - m_1^2 m_2^2 =
 \lambda(s_{12}, m_1^2, m_2^2) / 4$ is the flux factor;
 $p, n$ are the four-momenta of the proton-target and intra-deuteron
 nucleon, respectively;
 $\rho_{pN}^\pi$ is the relativistic invariant inclusive spectrum of
 pions arising from interaction of the intra-deuteron nucleon with the
 target proton.
 The functions $\Phi^{(u)}(|\bm q|)$ and  $\Phi^{(t)}(|\bm q|)$ depend on
 the relative momentum $q = n - D/2$ and contain full information
 about the structure of deuteron with one on-shell nucleon
 \cite{Illarionov:2002:EPJA}.

\begin{figure}
\begin{minipage}{\linewidth}
\begin{center}
  \setlength{\unitlength}{0.85mm}
\begin{picture}(100,90)   
 \put(0,0){
  \centering\epsfig{file=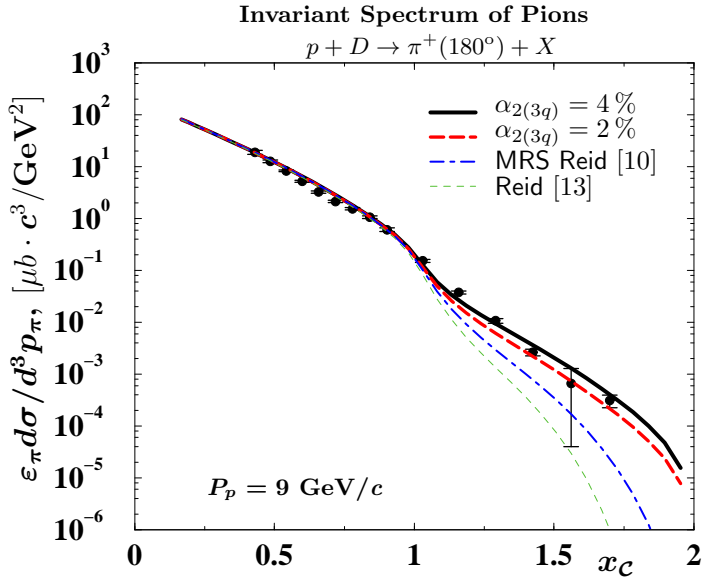, width=100\unitlength}}
 \put(29,86){\bfseries
   Invariant Spectrum of Pions}
 \put(39.5,81){\bfseries
   ${p + D \to \pi^+(180^\mathrm{o}) + X}$}
 \put(24,12){\bfseries
   $\boldsymbol{P_p = 9}$~GeV/\textit{c}}
 \put(85,0){\bfseries\large
  $\boldsymbol{x_{\mathcal C}}$}
 \put(-7,14){\bfseries\large
  \begin{sideways}
  $\boldsymbol{\varepsilon_\pi d\sigma/d^3p_\pi}$,
  $[\mu b \cdot \textit{c}^3/\text{GeV}^2]$
  \end{sideways}}
  \put(69,72){\normalsize
    $\alpha_{2(3q)} = 4\,\%$}
  \put(69,68){\normalsize
    $\alpha_{2(3q)} = 2\,\%$}
  \put(69,63){\normalsize
    \textsf{MRS Reid} \cite{Frankfurt:1981}}
  \put(69,59){\normalsize
    \textsf{Reid} \cite{Reid:1968}}
\end{picture}
\end{center}
\end{minipage}
\caption{
 The invariant pion spectrum calculated within the relativistic
 impulse approximation where  non-nucleonic components in the DWF
 \protect\cite{Lykasov:1993,Efremov:1988} have been included;
 its probability $\alpha_{2(3q)}$ is $0.02 \div 0.04$ (long-dashed and
 solid curves, respectively). One can have a good description of the data
 \protect\cite{Baldin:1985} for all $x_{\mathcal C}$.
}
\label{fig:Rho:6q}       
\end{figure}

 According to \cite{Frankfurt:1981}, large momenta of nucleons are due to
 few-nucleon correlations in the nucleus. Then the deuteron structure can
 be described by assuming quark degrees of freedom
 \cite{Lukyanov:1979,Burov:1984}.
 On the other hand, the shape of the high momentum tail of the nucleon
 distribution in the deuteron can be constructed on the basis of its true
 Regge asymptotic at $x \to 2$ \cite{Efremov:1988}, and the corresponding
 parameters can be found from the good description of the inclusive proton
 spectrum in the deuteron fragmentation $D p\to p X$ \cite{Efremov:1988}.
 According to \cite{Lykasov:1993,Efremov:1988}, one can
 write the following form for $\widetilde\Phi^{(u)}(|\vec q|)$:
\begin{equation}
 \Phi^{(u)}(|\vec q|) = \dfrac{E_{\vec k}/E_{\vec q}}{2(1 - x)}
 \widetilde\Phi^{(u)}(|\vec k|) \, .
\label{Phi:6q-(q->k)}
\end{equation}
where
\begin{align}
 \widetilde\Phi^{(u)}(|\vec k|) &= N_D^{-1} \dfrac{M_D^2}{m^2}
 \biggl[(1 - \alpha_{2(3q)}) \; \Bigl(U^2(|\vec k|) + W^2(|\vec k|)\Bigl)
\nonumber \\
 &+ \alpha_{2(3q)} \dfrac{8\pi x(1 - x)}{E_{\vec k}}
 \; G_{2(3q)}(x, \vec k_\bot) \biggl] \, ,
\label{Phi:6q}
\end{align}
 where $(x, \vec k_\bot)$ are the light-cone variables \cite{Frankfurt:1981}
 and $\vec k^2 = (m^2 + \vec k_\bot^2)/(4x(1-x)) - m^2$.
 The normalization factor $N_D^{-1}=\pi\sqrt{2/M_D}$ is chosen according to
 the non-re\-la\-ti\-vis\-tic normalization DWF \cite{Illarionov:2002:EPJA}.
 The parameter $\alpha_{2(3q)}$ is the probability for a non-nucleonic
 component in the deuteron which is a state of two colorless $(3q)$
 systems:
\begin{equation}
  G_{2(3q)}(x, \vec k_\bot) = \dfrac{b^2}{2\pi} \,
  \dfrac{\Gamma(A + B + 2)}{\Gamma(A + 1)\Gamma(B + 1)} \,
  x^A (1 - x)^B \, {\mbox{e}}^{-b |\vec k_\bot|}.
\label{6q}
\end{equation}
 Figure~\ref{fig:Rho:6q} presents the invariant pion spectrum calculated
 within the relativistic impulse approximation including the non-nucleonic
 component in the DWF \cite{Lykasov:1993,Efremov:1988}; its probability
 $\alpha_{2(3q)}$ is $0.02 \div 0.04$ (long-dashed and solid curves,
 respectively). One can see, that the inclusion of the non-nucleonic degrees
 of freedom within the approach suggested in
 \cite{Lykasov:1993,Efremov:1988}, the use of which has reproduced
 the data for the proton spectrum in the deuteron stripping, allows us also
 to describe the inclusive pion spectrum at all values of $x_{\mathcal C}$
 rather well (fig.~\ref{fig:Rho:6q}). However, the information contained in
 both observables is redundant, since it is the same deuteron properties
 that are the main ingredient in the analysis of both $Dp \to pX$ and
 $Dp \to \pi X$ reactions in the impulse approximation. Therefore, the
 calculation of the tensor analyzing power including the non-nucleonic
 degrees of freedom in fragmentation of the deuteron into pions can give
 us new independent information about the deuteron structure at small $N-N$
 distances and its comparison  with the data can be considered as a test of
 the modified DWF model used.
 Actually, in \cite{Efremov:1988} only a form of 
 $\widetilde\Phi^{(u)}(|\vec k|)$ has been constructed. However,
 to calculate $T_{20}$ it is not enough, the corresponding orbital waves
 have to be known. Let us assume that non-nucleonic degrees of freedom
 result in the main contribution to the ${\mathcal S}$ and ${\mathcal D}$
 waves of the deuteron wave function. Constructing new forms of these waves
 by including the non-nucleonic degrees of freedom we have to require
 that the square of the new DWF be equal to the one determined
 by eq.~(\ref{Phi:6q}). Introducing a mixing parameter $\alpha=\pi a/4$
 one can find the following forms of new ${\mathcal S}$ and ${\mathcal D}$
 waves
\begin{align}
  \widetilde{U}(|\vec k|) &=
   \sqrt{1-\alpha_{2(3q)}}U(|\vec k|) + \cos(\alpha) \Delta(|\vec k|) \, ;
\label{def:Unew} \\
  \widetilde{W}(|\vec k|) &=
   \sqrt{1-\alpha_{2(3q)}}W(|\vec k|) + \sin(\alpha) \Delta(|\vec k|) \, ,
\label{def:Wnew}
\end{align}
 where the function $\Delta(|\vec k|)$ has been obtained from the equation:
\begin{equation}
  \widetilde\Phi^{(u)}(|\vec k|) \ = \
  N_D^{-1}\dfrac{M_D^2}{m^2}
   \Bigl[\widetilde{U}^2(|\vec k|) + \widetilde{W}^2(|\vec k|)\Bigr] \, .
\label{rel:UW-Phi}
\end{equation}

\begin{figure}
\begin{minipage}{\linewidth}
\begin{center}
  \setlength{\unitlength}{0.85mm}
\begin{picture}(100,90)   
 \put(0,0){
  \centering\epsfig{file=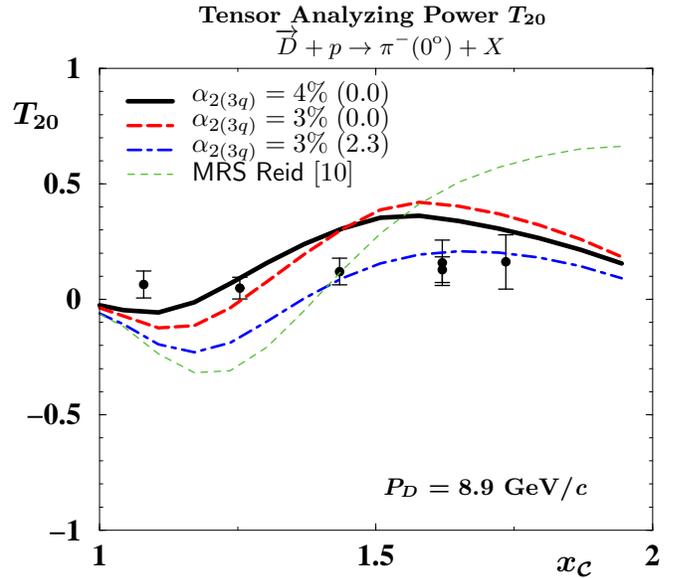, width=100\unitlength}}
 \put(29,86){\bfseries
   Tensor Analyzing Power $\boldsymbol{T_{20}}$}
 \put(41,81){\bfseries
   ${\overrightarrow{D} + p \to \pi^-(0^\mathrm{o}) + X}$}
 \put(58,12){\bfseries
   $\boldsymbol{P_D = 8.9}$~GeV/\textit{c}}
 \put(85,0){\bfseries\large
  $\boldsymbol{x_{\mathcal C}}$}
 \put(0,70){\bfseries\large
  $\boldsymbol{T_{20}}$}
  \put(28,73.7){\normalsize
    $\alpha_{2(3q)} = 4\% \; (0.0)$}
  \put(28,69.7){\normalsize
    $\alpha_{2(3q)} = 3\% \; (0.0)$}
  \put(28,65.7){\normalsize
    $\alpha_{2(3q)} = 3\% \; (2.3)$}
  \put(28,61.2){\normalsize
    \textsf{MRS Reid} \cite{Frankfurt:1981}}
\end{picture}
\end{center}
\end{minipage}
\caption{
 The tensor analyzing power $T_{20}$ calculated within the relativistic
 impulse approximation allowing for  non-nucleonic components in the DWF.
 The solid and long-dashed lines represent the calculations with the mixing
 parameter $a = 0.0$ and the probability $\alpha_{2(3q)} = 4\,\%$, $3\,\%$,
 respectively.
 The dot-dashed line corresponds to the calculation with the mixing
 parameter $a = 2.3$, which gives the curve closest to the data
 at $x_{\mathcal C} \geq 1.5$.
 The thin dashed curve corresponds to the Reid DWF \cite{Reid:1968}
 obtained by the minimal relativistic scheme (MRS) \cite{Frankfurt:1981}.
}
\label{fig:T20:6q}       
\end{figure}

 Figure \ref{fig:T20:6q} presents the analyzing power $T_{20}$
 calculated by using the functions $\widetilde{U}, \widetilde{W}$
 including the non-nucleonic components in the DWF, according to
 \cite{Lykasov:1993,Efremov:1988}.
 It is evident from fig.~\ref{fig:T20:6q} that the inclusion of non-nucleonic
 components in the DWF improves the description of the data 
 for $T_{20}$ at $x_{\mathcal C} > 1$. The best description of the
 observable is obtained for the value $a=2.3$ of the parameter $a$ entering
 into eqs.~(\ref{def:Unew}),(\ref{def:Wnew}).

 Main results can be summarized as the following. Very interesting
 experimental data on $T_{20}$ \cite{Afanasev:1998} showing approximately 
 zero values at $x_{\mathcal C} \geq 1$  are not reproduced by a theoretical 
 calculus using even different kinds of the relativistic DWF
 \cite{Illarionov:2002:EPJA}. 
 This may indicate a possible existence of non-nucleonic degrees of 
 freedom or basically new mechanism of pion production in the 
 kinematic region forbidden for free $N-N$ scattering.

 The inclusion of the non-nucleonic degrees of freedom within the approach 
 suggested in \cite{Lykasov:1993,Efremov:1988} allows us to describe
 experimental data about the inclusive pion spectrum at all the values of 
 $x_{\mathcal C}$ rather well, fig.~\ref{fig:Rho:6q}, and improve the
 description of data \cite{Afanasev:1998} concerning the analyzing power
 $T_{20}$ in the fragmentation of deuteron to pions, fig.~\ref{fig:T20:6q}.
 Of course, the inclusion of the non-nucleonic degrees of freedom
 in the analysis of $T_{20}$ is approximate, but can be considered as
 the indication of an important role of these degrees of freedom in studying
 polarization phenomena in the type of reactions considered.

 \textit{Acknowledgments.}
 A.Yu.I. is grateful to the Organizers of the International Conference on
 Quark Nuclear Physics 2002 for the kind invitation and financial support
 towards the participation.

%
%
%
%
%

\end{document}